# Structure dependent and strain tunable magnetic ordering in ultrathin chromium telluride

Jun Zhou[1,†], Xiaohe Song,[2,†] Jianwei Chai[1], Nancy Lai Mun Wong[1], Xiaoguang Xu[3], Yong Jiang[4], Yuan Ping Feng[2,5], Ming Yang[6,*], and Shijie Wang[1,*]

[1] *Institute of Materials Research & Engineering, A\*STAR (Agency for Science, Technology and Research), 2 Fusionopolis Way, Innovis, Singapore 138634, Singapore*

[2] *Integrative Sciences and Engineering Programme, NUS Graduate School, National University of Singapore, Singapore 117551, Singapore*

[3] *School of Materials Science and Engineering, University of Science and Technology Beijing, Beijing, 100083 China*

[4] *School of Electrical and Electronic Engineering, Tiangong University, Tianjin 300387, China*

[5] *Department of Physics, National University of Singapore, Singapore 117551, Singapore*

[6] *Department of Applied Physics, The Hong Kong Polytechnic University, Hung Hom, Kowloon, Hong Kong SAR, China*

[†] These authors contribute equally to this work.

[*] Correspondence should be addressed to M. Y. (mingyang@polyu.edu.hk) or S.J.W (sj-wang@imre.a-star.edu.sg)

Two-dimensional (2D) chromium tellurides have attracted considerable research interest for their high Curie temperatures. Their magnetic properties have been found diverse in various experiments, the understanding of which however remains limited. In this work, we report that the magnetic ordering of ultrathin chromium tellurides is structure dependent and can be tuned by external strain. Based on first-principles calculations and Monte Carlo simulations, we show long-range stable magnetism with high and low Curie temperature, and short-range magnetism in 2D $Cr_5Te_8$, $CrTe_2$, and $Cr_2Te_3$ layers, respectively. We further find that ferromagnetic-to-antiferromagnetic transition can be realized by 2% compressive strain for $CrTe_2$ and 2% tensile strain for $Cr_2Te_3$, and their magnetic easy axis is tuned from out-of-plane to in-plane by the medium tensile and compressive strain. This strain dependent magnetic coupling is found to be related to Cr-Cr direct exchange and the change of magnetic anisotropy is understood by the atom and orbital resolved magnetic anisotropy energy and second order perturbation theory. Our results reveal the important roles of the structure and strain in determining the magnetic ordering in 2D chromium telluride, shedding light on understanding of the diverse magnetic properties observed in experiments.

## 1. Introduction

Long-range magnetic ordering in low-dimension system was predicted to be unstable and can be destroyed by thermal fluctuation at any infinite temperature according to Mermin-Wagner theorem.[1] The breakthrough was achieved in 2017 by the experimental realization of truly two-dimensional (2D) magnet $CrI_3$ in monolayer limit and the underlying mechanism for the existence of 2D magnetism was uncovered by a scenario based on an opening of a spin-wave excitation gap caused by magnetic anisotropy.[2, 3] Consequently, more 2D magnetic materials such as $Fe_3GeTe_2$,[4] $FePS_3$,[5] $VSe_2$,[6] $MnSe_2$,[7] and $CrOCl$[8] have been reported. However, due to the decreased exchange coupling with thickness, the Curie temperature ($T_c$) for 2D magnetic materials is usually low. Although room-temperature ferromagnetism (FM) was first reported in 2D $VSe_2$, controversial observations on the absence of ferromagnetism in this system followed.[6, 9-11] Later, the room-temperature magnetism in $VSe_2$ was argued to be not intrinsic and was caused by defects.[12, 13] An alternative way to achieve room-temperature 2D magnetism was experimentally demonstrated to be feasible by applying ionic gating on $Fe_3GeTe_2$ monolayer.[4] However, intrinsic room-temperature 2D magnets are vigorously pursued for practical next-generation nanodevice applications.

Recently, the class of non-layered chromium telluride $Cr_xTe_y$ has attracted increasing research interest for high temperature magnetism.[14-17] The structures of bulk chromium tellurides ($Cr_xTe_y$) can be considered as layered octahedral $CrTe_2$ (an analogy to 1T $MoS_2$) with different concentrations of intercalated Cr atoms between the layers. Their monolayer growth thickness has been reported using chemical vapour deposition (CVD) and/or molecular beam epitaxy (MBE).[16, 18-21] Interestingly, room-temperature magnetism has been shown to persist in monolayer limit in this system.[15] However, due to the multivalent nature of the Cr, a large variation of stable compositions are possible for $Cr_xTe_y$,[22], and in experiment, diverse magnetic properties have been reported,[9, 15, 20, 23] which deserve a systematic study on the magnetic properties of $Cr_xTe_y$ in monolayer limit.

Another interesting fact about the low-dimensional magnetism is the tunability by multiple degree of freedom, such as stacking, thickness, strain, pressure, doping, external electric field, etc.[24-29] Among them, strain engineering is of particular interest for 2D materials, because the atomic thickness and strong chemical bonds within layers make 2D materials mechanically flexible. For example, $MoS_2$ monolayer has been shown to withstand up to 20% biaxial strain by indenting with an atomic force microscope.[30]. More importantly, strain engineering has been shown to tune the magnetic properties of 2D materials effectively. The ferromagnetic-to-antiferromagnetic transition as well as change of magnetic anisotropy between out-of-plane and in-plane have been reported.[31-34] Due to the rich physics and appealing applications, the concept straintronics has been proposed.[35, 36]

In this work, we focus on three Te terminated and high-symmetry $Cr_xTe_y$ monolayers with 3 (CrTe2), 5 ($Cr_2Te_3$) and 7 ($Cr_5Te_8$) atomic layers. Via first-principles calculations, we show that both the magnetic configurations and magnetic anisotropy of the $Cr_xTe_y$ monolayers can be effectively tuned by biaxial in-plane strains. The change of magnetic anisotropy was studied by atom and orbital resolved magnetic anisotropy energy (MAE) and understood by the second order perturbation theory. We show that spin-orbital coupling of Te atoms contribute most of the magnetic anisotropy energies of the $Cr_xTe_y$ monolayers and their overall magnetic anisotropies are the results of the competition between the positive contribution from $p_y/p_z$ hybridization and negative contribution from $p_x/p_y$ hybridizations. Our Monte Carlo simulations predict diverse magnetic properties of $Cr_xTe_y$ monolayers. $Cr_5Te_8$ has a Curie temperature as high as 440 K, but $CrTe_2$ monolayer has a much lower $T_c$, while long-range magnetism is absent in ML $Cr_2Te_3$ at finite temperature, in their strain-free pristine form.

## 2. Methodology

The first-principles calculations were performed based on the projector augmented-wave method as implemented in the Vienna ab initio simulation package (VASP.5.4.4). [37, 38] The exchange-correlation interaction was treated with the Perdew–Burke–Ernzerhof form of the generalized gradient approximation (GGA).[39] The Hubbard U of 3.7 eV was adopted to deal with the strongly correlated $d$ orbitals of Cr.[40] The plane-wave cutoff energy was set to 500 eV. The lattice parameters and atomic positions were fully relaxed until the energy and force on each atom were converged to less than $1 \times 10^{-5}$ eV and $1 \times 10^{-3}$ eV/Å, respectively. For magnetic anisotropy energy, the spin-orbital coupling was included with a higher accuracy of $1 \times 10^{-8}$ eV. Γ-centre $15 \times 15 \times 1$ and $31 \times 31 \times 1$ $k$-meshes were adopted for structure relaxation and static and MAE calculations, respectively.

The VAMPIRE package was used for metropolis Monte Carlo (MC) simulations [41], which was based on the following Heisenberg spin Hamiltonian:

$$\widetilde{H} = \sum_{ij} J_{ij} \mathbf{S}_i \cdot \mathbf{S}_j - A \sum_i S_{iz}^2 \tag{1}$$

where $\mathbf{S}_i$ and $\mathbf{S}_j$ are the classical Heisenberg spins with unit magnitude at the site $i$ and $j$. $J$ is the nearest neighbouring spin-spin coupling, and $A$ is the magnetic anisotropic energy. It is worth noting that we focus on the collinear exchange coupling in this work, the noncollinear terms in the Hamiltonian induced by spin-orbit coupling are not included. The simulated temperature dependent magnetization is fitted to the Curie-Bloch equation in the classical limit $m(T) = (1 - T/T_c)^\beta$, where β is the critical exponent.

The simulated spin systems for $CrTe_2$, $Cr_2Te_3$, $Cr_5Te_8$ consisted of 3600, 7200, 8000 spins, respectively, to ensure the converged magnetization-temperature curve. At each temperature, the orientations of spins were initially randomized, and all the spins were first thermalized for

10,000 equilibrium steps and then by 50,000 averaging steps to derive the thermal equilibrium magnetization.

## 3. Results and discussion

The top and side view of the structures of CrTe$_2$, Cr$_2$Te$_3$ and Cr$_5$Te$_8$ are shown in Fig. 1. The Cr atoms are coordinated with six surrounding Te atoms in the form of octahedrons. The octahedral crystal field split the five-degenerate $d$ orbitals of Cr to upper $e_g$ ($d_{z^2}$, $d_{x^2+y^2}$) and lower $t_{2g}$ ($d_{xy}$, $d_{xz}$, $d_{yz}$) states. The CrTe$_2$ monolayer has a 1T-phase structure with a space group of $P\bar{3}m1$. The relaxed in-plane lattice parameters are $a = b = 3.72$ Å, in line with previous reports.[42] The Cr$_2$Te$_3$ monolayer has a mirror symmetry along $c$ axis with a space group of $P\bar{6}m2$. The relaxed lattice parameters $a = b = 4.01$ Å are slightly larger than that of CrTe$_2$ monolayer. The Cr atoms form slightly longer bonds with the middle Te atoms (2.88 Å) than with the top and bottom Te atoms (2.76 Å). The Cr$_5$Te$_8$ monolayer can be taken as a CrTe$_2$ bilayer with interlayer intercalated Cr atoms in a concentration of 50%.[43] Such intercalation leads to a 2 × 1 periodic distortion along the $b$ axis of the top and bottom CrTe$_2$ sublayers. Both the upper and lower CrTe$_2$ sublayers resemble the T' phase, with uneven Cr-Cr distances of 3.86 Å and 4.12 Å in the left and right half of the unit cell, respectively. Overall, the Cr$_5$Te$_8$ monolayer has a space group of $P2/m$ with lattice parameters $a = 3.99$ Å and $b = 6.90$ Å.

The three Cr$_x$Te$_y$ monolayers studied in this work are magnetic conductors. The magnetic moments of Cr atoms are 3.47 $\mu_B$ in CrTe$_2$, 3.62 $\mu_B$ in Cr$_2$Te$_3$, and 3.56 $\mu_B$ (in top and bottom layer) and 3.72 $\mu_B$ (middle layer) for Cr$_5$Te$_8$, respectively. To study the magnetic ground state, we consider two antiferromagnetic (AFM) configurations in this work. Within each Cr atomic layer, the Cr atoms form a triangular lattice, in which a stripe AFM was adopted [denoted as in-plane antiferromagnetic (AFM$_{//}$)]. An example of the spin density for AFM$_{//}$ is shown in the inset of Fig. 2(a). On the contrary, the out-of-plane antiferromagnetic (AFM$_\perp$) configuration is the interlayer AFM between Cr atomic layers. Inset of Fig. 2(b) shows an example of the spin density for AFM$_\perp$. Their respective energy differences with the FM state are defined as $\Delta E_{//} = E_{\text{AFM}//} - E_{\text{FM}}$ and $\Delta E_\perp = E_{\text{AFM}\perp} - E_{\text{FM}}$.

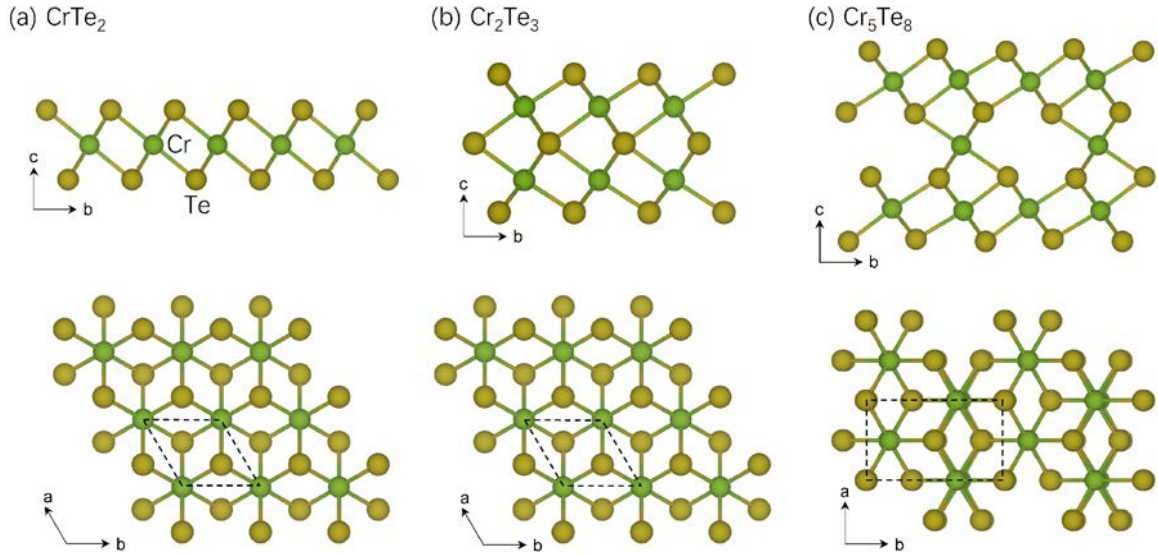

**Fig. 1.** Structural guidance in side (upper panel) and top (lower panel) view of (a) CrTe$_2$, (b) Cr$_2$Te$_3$, and (c) Cr$_5$Te$_8$ monolayers. The rhombus and rectangle in dash line indicate the unit cell in-plane lattice.

As the strain dependent energy difference shown in Fig. 2 (a)-(c), both $\Delta E_{//}$ and $\Delta E_\perp$ are positive without strain, indicating CrTe$_2$, Cr$_2$Te$_3$, and Cr$_5$Te$_8$ monolayers all prefer a FM configuration in the pristine form. It also shows that an in-plane biaxial strain effectively change both $\Delta E_{//}$ and $\Delta E_\perp$, in a monotonically proportional manner. A stark contrast is that $\Delta E_{//}$ increases with the in-plane lattice parameters while an opposite trend is observed for $\Delta E_\perp$. However, such contrast can be compatible from the perspective of Cr-Cr distance. As the in-plane Cr-Cr distance increases from compressive to tensile strain, the out-of-plane Cr-Cr decreases correspondingly. For example, the interlayer Cr-Cr distance in Cr$_2$Te$_3$ monolayer shrinks from 3.56 Å to 3.29 Å when the strain changes from 4% compressive to 4% tensile strain, respectively. Overall, the energy difference between AFM and FM increases with the Cr-Cr distance. A similar trend between exchange coupling strength and inter-atom distance can be found in magnetic metals, which is referred to as Bethe–Slater curve.[44, 45] Interestingly, a FM-to-AFM transition can be achieved by a 2% compressive strain for CrTe$_2$ monolayer and a 2% tensile strain for Cr$_2$Te$_3$. Considering monotonic nature of the curves, the FM-to-AFM transition may also be possible for Cr$_5$Te$_8$ but requires a much higher biaxial strain, beyond the range studied in this work.

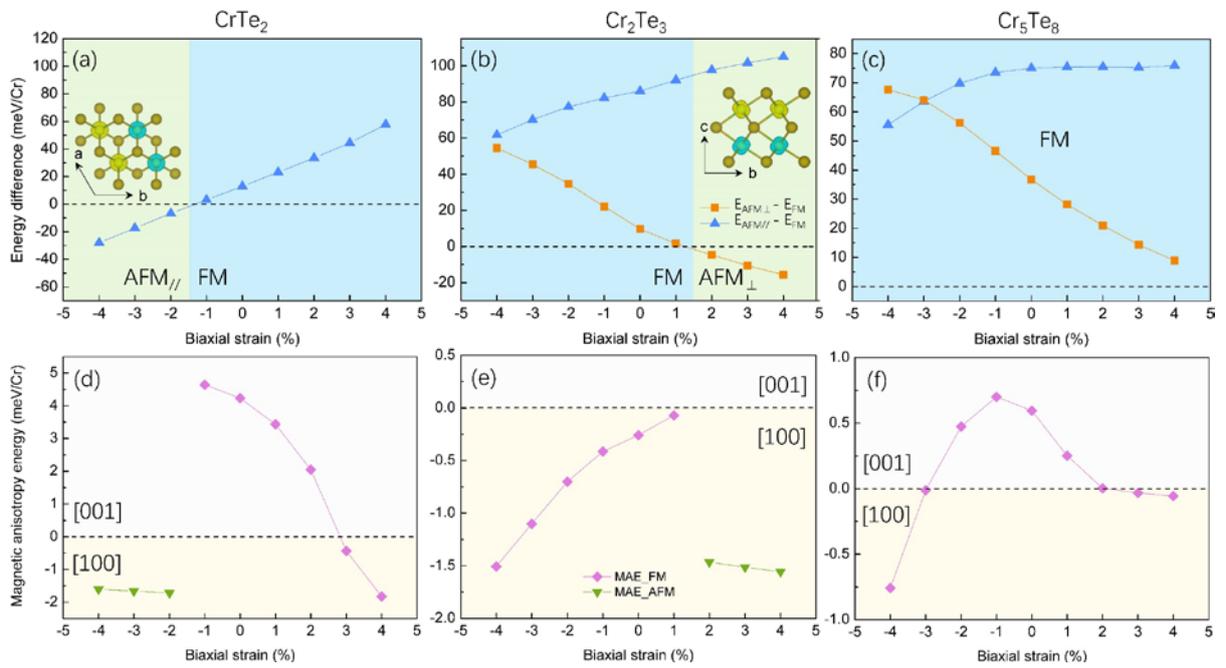

**Fig. 2**. Strain dependent energy differences between AFM and FM configurations for monolayer (a) CrTe$_2$, (b) Cr$_2$Te$_3$ and (c) Cr$_5$Te$_8$ as well as magnetic anisotropy energies for monolayer (d) CrTe$_2$, (e) Cr$_2$Te$_3$ and (f) Cr$_5$Te$_8$. The inset in (a) is the spin density of the in-plane antiferromagnetic configuration (AFM$_{//}$) and the one in (b) is out-of-plane antiferromagnetic configuration (AFM$_\perp$). The yellow and cyan colours represent the spin density of different spin polarization.

The magnetic anisotropy is the key for the long-range magnetic ordering in 2D magnetic materials. The strain effects on magnetic anisotropy of the three Cr$_x$Te$_y$ monolayers are studied by calculating the magnetic anisotropy energy (MAE). The magnetic anisotropy energy is defined by MAE = $E_{[100]} - E_{[001]}$, where the $E_{[100]}$ and $E_{[001]}$ are the total energies of the system in magnetic ground state when all the spins aligning along lattice vector $a$ and $c$, respectively. Thus, a positive value indicates an out-of-plane magnetic easy axis while a negative value suggests an in-plane anisotropy.

As the strain dependent MAE shown in the in Fig. 2 (d)-(f), both $CrTe_2$ and $Cr_5Te_8$ prefer an out-of-plane anisotropy with a value of 4.23 meV/Cr and 0.60 meV/Cr, respectively, while $Cr_2Te_3$ has an in-plane anisotropy with -0.26 meV/Cr. In addition, the MAE is quite sensitive with the strain effects for the FM ground state but changes less significantly for the AFM configuration. The sudden jumps of MAE in Fig. 2 (d) and (e) are the results of the different magnetic configurations (purple squares for FM and green triangles for AFM). Overall, the MAE is largest in the vicinity of the pristine case and decreases with both the compressive and tensile strain. More specifically, for $CrTe_2$, the biaxial strain can trigger an out-of-plane to in-plane magnetic anisotropy transition at 3% tensile strain and 2% compressive strain, the latter of which accompanies a ferromagnetic-to-antiferromagnetic transition simultaneously. For $Cr_2Te_3$, the magnetic anisotropy stays within the plane for the strain range studied in this work and is unlikely to change the magnetic easy axis in a wider range as the MAE keeps decreasing with larger strain. Thus, a long-range magnetic order is not expected in $Cr_2Te_3$ monolayer. The out-of-plane magnetic easy axis of $Cr_5Te_8$ is well maintained at the strain range from -2% to 1% and beyond this range, it turns to an in-plane anisotropy. These results suggest the destructive role of a large biaxial strain to the long-range FM ordering of $CrTe_2$, and $Cr_5Te_8$.

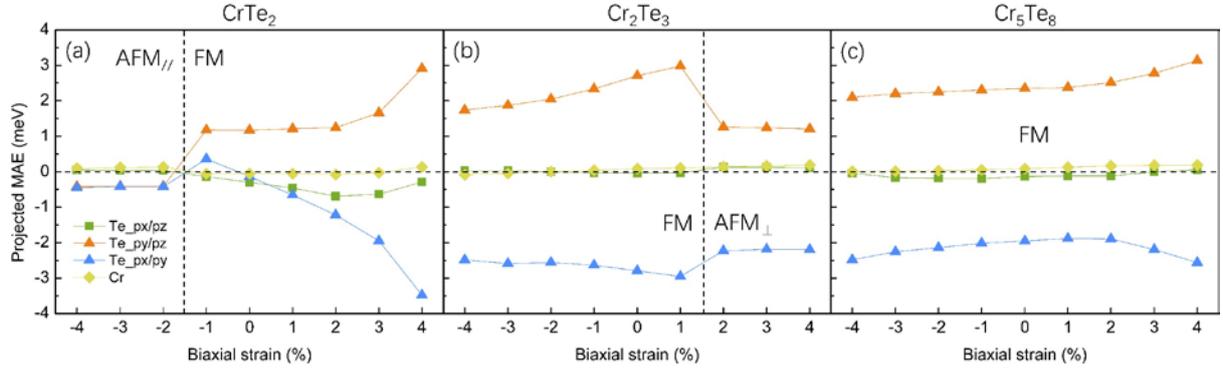

**Fig. 3.** Strain dependent atom and orbital resolved magnetic anisotropy energies for monolayer (a) $CrTe_2$, (b) $Cr_2Te_3$ and (c) $Cr_5Te_8$.

To understand the strain dependent magnetic anisotropy, we further analyse the atom and orbital resolved MAE for the $CrTe_2$, $Cr_2Te_3$ and $Cr_5Te_8$. As shown in Fig. 3, the main contribution of MAE is from Te atoms. It is expected as the spin-orbital coupling (SOC) strength increases geometrically with the atomic number $Z$ ($Z_{Cr} = 24$ and $Z_{Te} = 52$).[46] Thus, we focus on the orbital resolved MAE of Te in the following discussions. According to the second order perturbation theory, the SOC coupling is non-zero only by the hybridization of different orbitals.[47] Within the $5p$ orbital of Te, the MAE from the hybridization of $p_x/p_z$ is also negligible. And the overall MAE is a result of the competition between the positive (out-of-plane) $p_y/p_z$ hybridization and the negative (in-plane) $p_x/p_y$ hybridization. Besides, both the $p_y/p_z$ and $p_x/p_y$ hybridizations are less sensitive to the strain in the AFM configuration than the FM state, leading the less strain dependent overall MAE of the AFM configuration.

The different sign and strength of the orbital resolved MAE of Te are further understood based on the second order perturbation theory, according to which the MAE from the orbital hybridization can be determined by the following equation.[47]

$$\Delta E = E^{\sigma\sigma'}(x) - E^{\sigma\sigma'}(z)$$

$$= (2\delta_{\sigma\sigma'} - 1)\xi^2 \sum_{o^\sigma, u^{\sigma'}} \frac{\left|\langle o^\sigma|L_z|u^{\sigma'}\rangle\right|^2 - \left|\langle o^\sigma|L_x|u^{\sigma'}\rangle\right|^2}{E_u^{\sigma'} - E_o^\sigma} \quad (2)$$

where, ξ is the SOC constant; $o$ and $u$ indicate the occupied and unoccupied states, respectively; $\sigma$ and $\sigma'$ are the different spin states.

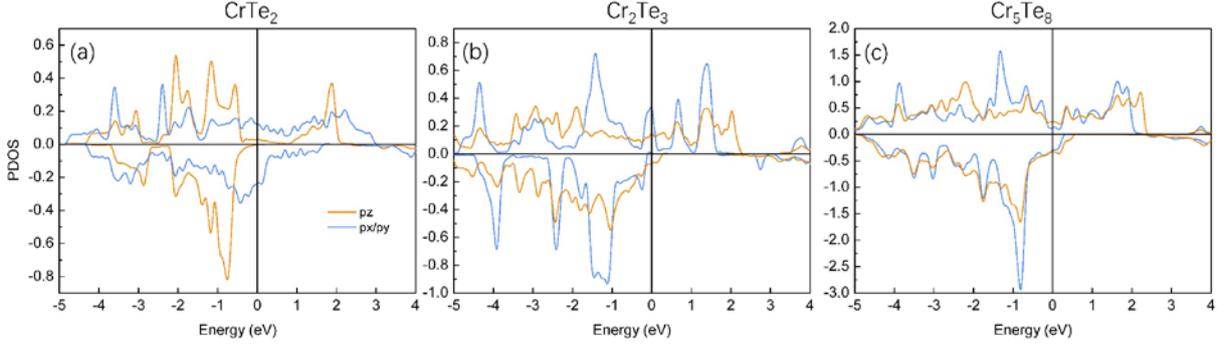

**Fig. 4.** Projected density of states for Te $p$ orbitals for monolayer (a) CrTe$_2$, (b) Cr$_2$Te$_3$ and (c) Cr$_5$Te$_8$.

The projected density of states (PDOS) of Te $5p$ orbitals for CrTe$_2$, Cr$_2$Te$_3$ and Cr$_5$Te$_8$ monolayers are shown in Fig. 4. Near the Fermi level, the unoccupied states are mainly in the spin-up channel ($|u^\uparrow\rangle$) while the spin-down states contribute more to the occupied states ($|o^\downarrow\rangle$). Thus, the MAE from contributions of hybridization of $5p$ orbitals are:

$$\Delta E(p_x, p_y) = (-1) \frac{\left|\langle p_x^\downarrow|L_z|p_y^\uparrow\rangle\right|^2 - \left|\langle p_x^\downarrow|L_x|p_y^\uparrow\rangle\right|^2}{E_u^\uparrow - E_o^\downarrow} = \frac{-1}{E_u^\uparrow(p_x) - E_o^\downarrow(p_y)}$$

$$\Delta E(p_x, p_z) = (-1) \frac{\left|\langle p_x^\downarrow|L_z|p_z^\uparrow\rangle\right|^2 - \left|\langle p_x^\downarrow|L_x|p_z^\uparrow\rangle\right|^2}{E_u^\uparrow - E_o^\downarrow} = 0$$

$$\Delta E(p_y, p_z) = (-1) \frac{\left|\langle p_y^\downarrow|L_z|p_z^\uparrow\rangle\right|^2 - \left|\langle p_y^\downarrow|L_x|p_z^\uparrow\rangle\right|^2}{E_u^\uparrow - E_o^\downarrow} = \frac{1}{E_u^\uparrow(p_y) - E_o^\downarrow(p_z)}$$

These results are in line with the negative (in-plane), negligible and positive (out-of-plane) contribution from the $p_x/p_y$, $p_x/p_z$ and $p_y/p_z$ hybridization to the orbital resolved MAE, respectively.

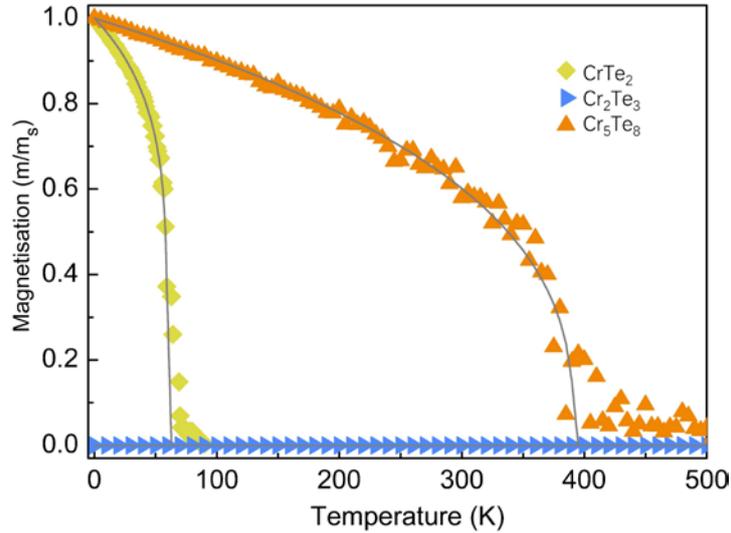

**Fig. 5.** The temperature dependence of normalized magnetization for monolayer $CrTe_2$, $Cr_2Te_3$ and $Cr_5Te_8$.

At last, we perform Monte Carlo simulations based on the Heisenberg model to study the temperature dependent magnetization for $CrTe_2$, $Cr_2Te_3$ and $Cr_5Te_8$ monolayers. As shown in Fig. 5, both $CrTe_2$ and $Cr_5Te_8$ monolayers can form a long-range ferromagnetic ordering due to their out-of-plane easy magnetic axis, leading to a non-zero overall magnetization. However, the $Cr_5Te_8$ monolayer has a much larger Curie temperature (around 391 K) than that of $CrTe_2$ (59 K). On the contrary, $Cr_2Te_3$ monolayer has an in-plane magnetic anisotropy, and its long-range magnetic ordering is disturbed by any amount of thermal energy. Thus, a macroscopic magnetization is missing at any temperature for $Cr_2Te_3$ monolayer.

We note that a Curie temperature of around 200 K has been shown experimentally for $CrTe_2$ monolayer.[14] The discrepancy may rise from different aspects. Firstly, the magnetic exchange strength used in MC simulations is extracted from an energy mapping of the AFM and FM configurations. However, the metallic nature of CrTe$_2$ suggests the importance of electronic excitations in determining the total energy beside spins. And the Heisenberg spin model is classical without considering the quantum nature of spins. These might result in an underestimated Curie temperature. Secondly, the $T_C$ of CrTe$_2$ was estimated experimentally by element-specific XMCD characterization due to the week signal of magnet in monolayer limit, which however in a strict sense only probes local magnetic moments.[14] Thirdly, the CrTe$_2$ monolayer was grown on a substrate in the experiment, which may introduce strain, doping and interface effects. It is interesting to note the $\Delta E_{//}$ of CrTe$_2$ increases from 12.96 meV/Cr to 23.15 meV/Cr by only 1% strain, indicating a much higher $T_C$ for CrTe$_2$ under strain. Further studies are needed to clarify this discrepancy.

In conclusion, we have systematically studied the strain effects on the magnetic configurations and magnetic anisotropy of Te-terminated high-symmetry Cr$_x$Te$_y$ with 3, 5 and 7 atomic layers. Our results show that the biaxial strain can effectively tune the magnetic configurations as well as the magnetic anisotropy of these three monolayers. The strain dependent magnetic ground state is understood by the accordingly changed Cr-Cr distances. And the atom and orbital resolved MAE and the second order perturbation theory suggest the strain tuneable magnetic anisotropy is the result of the competition between the negative contribution from Te $p_x$/$p_y$ hybridization and the positive $p_y$/$p_z$ contribution. The Monte Carlo simulations reproduce the high Curie temperature for Cr$_5$Te$_8$ monolayer as well as a low Tc and absence of long-range magnetic ordering for CrTe$_2$ and Cr$_2$Te$_3$, respectively. Our results may shed lights on the understanding of the experimental observations of this hot 2D system and the strain tuneable magnetic properties are appealing for spintronic and straintronic applications.

**Author Contributions**

**Conflicts of interest**

There are no conflicts to declare.

**Acknowledgements**

This work is supported by the Ministry of Education, Singapore, under its MOE Tier 1 Awards R-144-000-441-114 & R-144-000-413-114. M.Y. acknowledges the funding support (project ID: 1-BE47) from The Hong Kong Polytechnic University. We acknowledge Centre for Advanced 2D Materials and Graphene Research at National University of Singapore, and the National Supercomputing Centre of Singapore for providing computing resources.